\documentclass[12pt]{article}

\usepackage{latexsym,amssymb,amsfonts,amsmath}
\usepackage[dvips]{graphicx}
\usepackage{setspace}

\newcommand{\tr}{{\rm Tr}}
\newcommand{\de}{\partial}
\newcommand{\dde}[1]{\partial_{#1}}
\newcommand{\la}{\langle}
\newcommand{\ra}{\rangle}

\newcommand{\diag}{{\rm diag}}
\newcommand{\sign}{{\rm sign}}
\newcommand{\atan}{\tan^{-1}}
\newcommand{\GeV}{{\rm GeV}}

\newcommand{\qbar}{\bar{q}}

\renewcommand{\Re}{{\rm Re\,}}

\renewcommand{\thefootnote}{\fnsymbol{footnote}}

\begin{document}

\noindent \hspace{1cm}  \hfill IFUP--TH/2009--3 \hspace{1cm}\\
\mbox{}                 \hfill February 2009 \hspace{1cm}\\

\begin{center}
  \vspace*{1.0cm}
  \begin{doublespace}
    {\Large 
      {A NONPERTURBATIVE FOUNDATION OF THE EUCLIDEAN--MINKOWSKIAN DUALITY OF
      WILSON--LOOP CORRELATION FUNCTIONS}}
  \end{doublespace}
  \vspace*{0.5cm}
  {\large Matteo Giordano\footnote{E--mail: matteo.giordano@df.unipi.it}
    and Enrico Meggiolaro\footnote{E--mail: enrico.meggiolaro@df.unipi.it} }\\
  \vspace*{0.5cm}{\normalsize
    {Dipartimento di Fisica, Universit\`a di Pisa,\\
      and INFN, Sezione di Pisa,\\
      Largo Pontecorvo 3,
      I--56127 Pisa, Italy.}}\\
\end{center}
\vspace*{\stretch{1}}

%\maketitle

\setcounter{footnote}{0}
\renewcommand{\thefootnote}{\arabic{footnote}}
\thispagestyle{empty}

\abstract{In this letter we discuss the analyticity properties of the
  Wilson--loop correlation functions relevant to the problem of {\it
    soft} high--energy scattering, directly at the level of the
  functional integral, in a genuinely nonperturbative way. The
  strategy is to start from the Euclidean theory and to
  push the dependence on the relevant variables $\theta$ (the relative
  angle between the  loops) and $T$ (the half--length of the loops)
  into the action by means of a field and coordinate transformation,
  and then to allow them to take complex values. 
  In particular, we determine the analyticity domain of the relevant
  Euclidean correlation function, and we show that the corresponding
  Minkowskian quantity is recovered with the usual double analytic
  continuation in $\theta$ and $T$ inside this domain. The formal
  manipulations of the functional integral are justified making use of
  a lattice regularisation. The new {\it rescaled} action so derived
  could also be used directly to get new insights (from first
  principles) in the problem of {\it soft} high--energy scattering.}

\section{Introduction}

In recent years, starting from the seminal paper~\cite{Nachtmann91} by
O.~Nachtmann,
the long--standing problem of {\it soft} high--energy scattering in
strong interactions has been approached in the framework of
nonperturbative QCD, with functional--integral techniques (for a review
see Refs.~\cite{Dosch,pomeron-book}). 
In this approach the parton--parton elastic scattering amplitudes, 
at high center--of--mass energy $\sqrt{s} \gg 1 \GeV$
and small transferred momentum $\displaystyle\sqrt{|t|} \lesssim 1 \GeV$, 
are described by certain (properly normalised) correlation functions of 
two infinite lightlike Wilson lines, running along the classical trajectories 
of the colliding
partons~\cite{Nachtmann91,Verlinde,Korchemsky,Meggiolaro96,Meggiolaro01}.   
These correlation functions suffer from infrared (IR)
divergences~\cite{Verlinde,Korchemsky},  
which can be regularised considering  Wilson lines of finite length $2T$ along
the classical trajectories of partons with non--zero mass $m$, so forming
a finite hyperbolic angle
$\chi{\simeq}\log(s/m^2)$ (for ${s\to\infty}$) in Minkowski 
space--time~\cite{Verlinde,Korchemsky,Meggiolaro97,Meggiolaro98,Meggiolaro02}. 
IR divergences can be avoided from the onset, considering instead the
elastic scattering amplitude of two colourless states, e.g. two
$q\qbar$ meson states, which is expected to be an IR--finite
quantity~\cite{BL}. It has been shown 
in~\cite{DFK,Nachtmann97,BN,LLCM1} (see
also~\cite{Dosch,pomeron-book}) that in this case the meson--meson
elastic scattering amplitudes can be reconstructed from the
correlation functions of two Wilson loops (which describe the
scattering of two colour dipoles of fixed transverse size) running along
the trajectories of the colliding hadrons, after folding them with
appropriate wave functions describing the interacting mesons. In this
letter we will be concerned with meson--meson elastic scattering only,
and thus only with Wilson--loop correlation functions.  

It has been shown
in~\cite{Meggiolaro97,Meggiolaro98,Meggiolaro02,Meggiolaro05,crossing,Meggiolaro07}
that, under certain analyticity hypotheses, the relevant correlation
functions can be reconstructed from the ``corresponding'' correlation
functions of two Euclidean Wilson lines or Wilson loops, of finite
length $2T$, and forming an angle $\theta$ in Euclidean space, by
means of the double analytic continuation $\theta \to -i\chi$, $T\to
iT$. This ``Euclidean--Minkowskian duality'' of Wilson--line/loop
correlation functions has made possible to approach the problem of
{\it soft} high--energy scattering with the nonperturbative techniques
of Quantum Field Theory, usually available only in Euclidean space,
such as the {\it Instanton Liquid Model}~\cite{instanton1}, the
{\it Stochastic Vacuum Model}~\cite{LLCM2}, the {\it AdS/CFT
correspondence}~\cite{JP1,JP2}, and, recently, {\it Lattice Gauge 
Theory}~\cite{lattice}. We stress the fact that these
relations have been explicitly verified in perturbation
theory~\cite{Meggiolaro97,Meggiolaro05,crossing,BB}, up to $O(g^6)$
in the loop--loop case~\cite{BB}, while a nonperturbative
justification of the underlying analyticity hypotheses was still
lacking up to now, except in the case of {\it quenched} QED where an exact 
calculation can be performed both in the Euclidean and in the Minkowskian
theories~\cite{Meggiolaro05}.  
Since the analytic--continuation relations are expected to be 
{\it exact}, i.e., to hold beyond perturbation theory, it is 
important to provide them with a genuinely nonperturbative
foundation, and this is the purpose of this work.

In this letter we approach the analyticity issues related to the case
of meson--meson scattering directly at the level of the functional
integral. The strategy is to push the dependence on the relevant
variables into the action by means of a field and coordinate
transformation, and then to allow them to take complex values. 
In particular, we determine the analyticity domain of the relevant
Euclidean correlation function, and we show that the corresponding
Minkowskian quantity is recovered with the usual double analytic
continuation in $\theta$ and $T$ inside this domain; moreover, the
extra conditions that allow one to derive the {\it crossing--symmetry
  relations} found in Ref.~\cite{crossing} are shown to be
satisfied. The formal manipulations of the functional integral used to
obtain these results are justified making use of a lattice
regularisation. 

Due to the IR--finiteness of the scattering amplitude of two
colourless states in gauge theory, already mentioned above, 
one expects that the Wilson--loop correlation functions relevant to
the meson--meson case are finite in the limit $T\to\infty$. 
It has been shown in~\cite{Meggiolaro05} that in this case, after the
removal of the IR cutoff $T$, the two quantities are 
still connected by the same analytic continuation in the angular
variable only, as long as some requirements on the
correlators as functions of the complex variable $T$ are met. In this
letter we give more refined arguments supporting this conclusion.

The outline of this letter is as follows. In section 2 we briefly
recall the definitions of the relevant quantities, and in section 3 we
perform a field and coordinate transformation that pushes the whole
dependence on the relevant variables into the action. In section 4 we
discuss the analyticity properties of the correlation functions in the
pure--gauge theory case, using a lattice regularisation to justify the
formal manipulations of the functional integral; at the end of section
4 we also briefly discuss the inclusion of fermions. Some concluding
remarks and prospects for the future are shown in section 5.

\section{High--energy meson--meson scattering and Wilson--loop
  correlation functions} 

In this section we briefly recall, for the benefit of the reader, the
main points of the functional--integral  approach to the problem of elastic
meson--meson scattering. We refer the interested reader to the original
papers~\cite{DFK,Nachtmann97,BN,LLCM1}. We shall use the same notation
adopted in~\cite{lattice}, where a more detailed presentation can be
found. 

The elastic scattering amplitudes of two mesons (taken for simplicity
with the same mass $m$) in the {\it soft} high--energy regime can be
reconstructed in two steps. One
first evaluates the scattering amplitude 
of two $q\qbar$ colour dipoles of fixed transverse sizes $\vec{R}_{1\perp}$
and $\vec{R}_{2\perp}$, and fixed longitudinal momentum fractions $f_1$ and
$f_2$ of the two quarks in the two dipoles, respectively; the mesonic
amplitudes are then obtained after folding the dipoles' amplitudes
with the appropriate squared wave functions, describing the
interacting mesons. The dipole--dipole amplitudes are given by the
2--dimensional Fourier transform, with respect to the transverse
distance $\vec{z}_{\perp}$, of the normalised (connected) correlation
function of two rectangular Wilson loops,
\begin{equation}
  {\cal M}_{(dd)} (s,t;\vec{R}_{1\perp},f_1,\vec{R}_{2\perp},f_2) \equiv
-i~2s \displaystyle\int d^2 \vec{z}_\perp
e^{i \vec{q}_\perp \cdot \vec{z}_\perp}
{\cal C}_M(\chi;\vec{z}_\perp;1,2) ,
\label{scatt-loop}
\end{equation}
where the arguments ``$1$'' and ``$2$'' stand for ``$\vec{R}_{1\perp},
f_1$'' and ``$\vec{R}_{2\perp}, f_2$'' respectively, $t =
-|\vec{q}_\perp|^2$ ($\vec{q}_\perp$ being the transferred momentum)
and $s=2m^2(1+\cosh\chi)$.
The correlation function ${\cal C}_M$ is defined as the
limit $\displaystyle {\cal C}_M \equiv \lim_{T\to\infty} {\cal G}_M $
of the correlation function of two loops of finite length $2T$,
\begin{equation}
  {\cal G}_M(\chi;T;\vec{z}_\perp;1,2) \equiv
{ \langle {\cal W}^{(T)}_1 {\cal W}^{(T)}_2 \rangle \over
\langle {\cal W}^{(T)}_1 \rangle
\langle {\cal W}^{(T)}_2 \rangle } - 1,
\label{GM}
\end{equation}
where $\langle\ldots\rangle$ are averages in the sense of the QCD
functional integral, and
\begin{equation}
{\cal W}^{(T)}_{1,2} \equiv
{\dfrac{1}{N_c}} \tr \left\{ {\cal P} \exp
\left[ -ig \displaystyle\oint_{{\cal C}_{1,2}} A_{\mu}(x) dx^{\mu} \right]
\right\} 
\label{QCDloops}
\end{equation}
are Wilson loops in the fundamental representation of $SU(N_c)$; the
paths are made up of the quarks and antiquarks classical trajectories,
\begin{align}
  {\cal C}_1 :& \quad X_{1q}^\mu(\tau) = z^\mu + {p_1^\mu \over m}
  \tau + (1-f_1) R_1^\mu, ~~~~ \nonumber\\ 
\phantom{{\cal C}_1 :}& \quad  X_{1\bar{q}}^\mu(\tau) = z^\mu + {p_1^\mu
  \over m} \tau - f_1 R_1^\mu  ,
\nonumber \\
{\cal C}_2 :& \quad X_{2q}^\mu(\tau) = {p_2^\mu \over m} \tau +
(1-f_2) R_2^\mu, ~~~~ \nonumber\\ 
\phantom{{\cal C}_2 :}& \quad X_{2\bar{q}}^\mu(\tau) = {p_2^\mu \over m} \tau - f_2 R_2^\mu ,
\label{traj}
\end{align}
with $\tau\in [-T,T]$, and closed by straight--line paths in the
transverse plane at $\tau=\pm T$. Here
\begin{eqnarray}
&p_1 =
m \left( \cosh {\chi \over 2},\sinh {\chi \over 2},\vec{0}_\perp \right) ,~~~
p_2 =
m \left( \cosh {\chi \over 2},-\sinh {\chi \over 2},\vec{0}_\perp \right),
\label{p1p2}
\end{eqnarray}
and moreover, $R_1 = (0,0,\vec{R}_{1\perp})$, $R_2 = (0,0,\vec{R}_{2\perp})$,
$z = (0,0,\vec{z}_\perp)$, and $f_i$
is the longitudinal momentum fraction of quark $i$, $f_i\in [0,1]$.

The Euclidean counterpart of Eq.~(\ref{GM}) is
\begin{equation}
  {\cal G}_E(\theta;T;\vec{z}_\perp;1,2) \equiv
\dfrac{ \langle \widetilde{\cal W}^{(T)}_1 \widetilde{\cal W}^{(T)}_2 \rangle_E}
{\langle \widetilde{\cal W}^{(T)}_1 \rangle_E
\langle \widetilde{\cal W}^{(T)}_2 \rangle_E } - 1,
\label{GE}
\end{equation}
where now $\langle\ldots\rangle_E$ is the average in the sense of the
Euclidean QCD functional integral, and the Euclidean Wilson loops
\begin{equation}
\widetilde{\cal W}^{(T)}_{1,2} \equiv
{\dfrac{1}{N_c}} \tr \left\{ {\cal P} \exp
\left[ -ig \displaystyle\oint_{\widetilde{\cal C}_{1,2}} A_{E\mu}(x_E) dx_{E\mu} \right]
\right\} 
\label{QCDloopsE}
\end{equation}
are calculated on the following straight--line paths\footnote{The
  fourth Euclidean coordinate $X_{E4}$ is taken to be the ``Euclidean
  time''.},
\begin{align}
  \widetilde{\cal C}_1 :& \quad
X^{1q}_{E\mu}(\tau) = z_{E\mu} + {p_{1E\mu} \over m} \tau
+ (1-f_1) R_{1E\mu}, ~~~~ \nonumber\\
\phantom{\widetilde{\cal C}_1 :}& \quad X^{1\bar{q}}_{E\mu}(\tau) =
z_{E\mu} + {p_{1E\mu} \over m} \tau  
- f_1 R_{1E\mu} ,
\nonumber \\
\widetilde{\cal C}_2 :& \quad
X^{2q}_{E\mu}(\tau) = {p_{2E\mu} \over m} \tau + (1-f_2) R_{2E\mu},
~~~~\nonumber\\ 
\phantom{\widetilde{\cal C}_2 :}& \quad X^{2\bar{q}}_{E\mu}(\tau) =
        {p_{2E\mu} \over m} \tau - f_2 R_{2E\mu} , 
\label{trajE}
\end{align}
with $\tau\in [-T,T]$, and closed by straight--line paths in the
transverse plane at $\tau=\pm T$. Here
\begin{eqnarray}
&p_{1E} =
m \left( \sin{\theta \over 2}, \vec{0}_\perp, \cos{\theta \over 2} \right) ,~~~
p_{2E} =
m \left( -\sin{\theta \over 2}, \vec{0}_\perp, \cos{\theta \over 2} \right),
\label{p1p2E}
\end{eqnarray}
and $R_{1E} = (0,\vec{R}_{1\perp},0)$, $R_{2E} =
(0,\vec{R}_{2\perp},0)$, $z_E = (0,\vec{z}_\perp,0)$ (the
transverse vectors are taken to be equal in the two
cases).
Again, we define the correlation function with the IR cutoff removed
as $\displaystyle {\cal C}_E \equiv \lim_{T\to\infty} {\cal G}_E $. 

It has been shown
in~\cite{Meggiolaro97,Meggiolaro98,Meggiolaro02,Meggiolaro05} that 
the correlation functions in the two theories are connected by the
{\it analytic--continuation relations}
\begin{eqnarray}
{\cal G}_M(\chi;T;\vec{z}_\perp;1,2)
&=& \overline{\cal G}_E (-i\chi;iT;\vec{z}_\perp;1,2) ,
\qquad \forall\chi\in {\cal I}_{M},\nonumber \\
  {\cal G}_E(\theta;T;\vec{z}_\perp;1,2)
&=& \overline{\cal G}_M (i\theta;-iT;\vec{z}_\perp;1,2) ,
\qquad \forall\theta\in {\cal I}_{E}.
\label{analytic}
\end{eqnarray}
Here we denote with an overbar the {\it analytic extensions} of the
Euclidean and Min\-kow\-skian correlation functions, starting from the
real intervals  ${\cal I}_E\equiv (0,\pi)$ and
${\cal I}_M \equiv (0,\infty)=\mathbb{R}^+$ of the respective angular variables,
with positive real $T$ in both cases, into domains of the complex
variables $\theta$ 
(resp.~$\chi$) and $T$ in a two--dimensional complex space; these
domains are assumed to contain the intervals  $-i{\cal I}_M$ (at
positive imaginary $T$)\footnote{We use here and in the following the
  notation $\alpha +   \beta{\cal I} =   \{ \alpha + \beta z | z\in
  {\cal I} \}$.} and  $i{\cal I}_E$ (at negative imaginary $T$) in 
the two cases, respectively. 
Under certain analyticity hypotheses in the $T$ variable, the
following relations are obtained for the correlation functions with
the IR cutoff $T$ removed~\cite{Meggiolaro05}:
\begin{eqnarray}
{\cal C}_M(\chi;\vec{z}_\perp;1,2) &=&
\overline{\cal C}_E(-i\chi;\vec{z}_\perp;1,2) ,
\qquad \forall\chi\in {\cal I}_M,\nonumber \\
  {\cal C}_E(\theta;\vec{z}_\perp;1,2) &=&
\overline{\cal C}_M(i\theta;\vec{z}_\perp;1,2) ,
\qquad \phantom{-}\forall\theta\in {\cal I}_E.
\label{analytic_C}
\end{eqnarray}
Finally, we recall the {\it crossing--symmetry
  relations}~\cite{crossing} 
\begin{eqnarray}
\overline{\mathcal{G}}_M(i\pi-\chi;T;\vec{z}_{\perp};1,2)
&=&\mathcal{G}_M(\chi;T;\vec{z}_{\perp};1,\overline{2}) \nonumber \\
&=&\mathcal{G}_M(\chi;T;\vec{z}_{\perp};\overline{1},2) ,
\quad \forall\chi\in {\cal I}_M,
\nonumber \\
  \mathcal{G}_E(\pi-\theta;T;\vec{z}_{\perp};1,2)
&=&\mathcal{G}_E(\theta;T;\vec{z}_{\perp};1,\overline{2}) \nonumber\\ 
&=&\mathcal{G}_E(\theta;T;\vec{z}_{\perp};\overline{1},2) ,
\quad\forall\theta\in {\cal I}_E ,
\label{eq:crossrel}
\end{eqnarray}
that hold for every positive real $T$, and thus also for the
correlation functions 
with the IR cutoff removed; here the arguments ``$\overline{1}$'' and
``$\overline{2}$'' stand for ``$-\vec{R}_{1\perp}, 1-f_1$'' and
``$-\vec{R}_{2\perp}, 1-f_2$'' respectively. The Euclidean relation in
(\ref{eq:crossrel}) holds without any analyticity hypothesis, while in
the Min\-kow\-skian case the analyticity domain for the analytic
extension $\overline{\cal G}_M$ should include also the interval (in the
complex--$\chi$ plane) ${\cal  I}_M^{(c)}= i\pi - {\cal I}_M$ (for
positive real $T$), where the physical amplitude for the  ``crossed''
channel is then expected to lie. 

A more precise formulation of the analytic--continuation relations will
be given in section 4, where we will determine the analyticity
domain of the Euclidean correlation function making use of
nonperturbative arguments directly at the level of the functional
integral. We will also show that the analyticity hypotheses
required for the validity of Eqs.~(\ref{analytic_C}) and
(\ref{eq:crossrel}) are satisfied.

\section{Field and coordinate transformation}

To address the issue of the analytic extension of the correlation
functions to complex values of the angular variables and of $T$, we
shall appropriately rescale the coordinates and fields, in order for
the dependence on the relevant variables to drop from the Wilson--loop
operators, and to move into the action. For the time being we
consider the pure--gauge theory only; the inclusion of fermions will
be briefly discussed at the end of this section.

We first rescale~\cite{Meggiolaro98,Meggiolaro02,Meggiolaro05}
$\tau\to \alpha\tau$ in the ${\cal P}$--exponentials 
corresponding to the longitudinal sides, so that the paths are
redefined to be the ones with $p_i/m$ and $p_{Ei}/m$ substituted by
$\alpha p_i/m$ and $\alpha p_{Ei}/m$ (and $\tau\in
[-T/\alpha,T/\alpha]$); we can set $\alpha= T/T_0$ with $T_0$ some
fixed time (length) scale, thus showing that the loops depend on $T$
only through the combinations\footnote{The ${\cal P}$--exponentials
  corresponding to the transverse sides explicitly depend on $T
  p_{Ei}/m = T_0(T/T_0) p_{Ei}/m$.} $(T/T_0) p_i/m$ and $(T/T_0)
p_{Ei}/m$.

Next, we rescale coordinates and fields as follows. To unify the
treatment of the Euclidean and Minkowskian cases we use the same
symbol $\phi_\mu$ for the transformed gauge fields, and $y^\mu$ for
the transformed coordinates (we can use upper indices for the new
coordinates also in the Euclidean case without ambiguity), with
$\mu=0,1,2,3$ (we identify 0 and 4 as indices in the Euclidean case). We then
set in the Minkowskian case
\begin{align}
  y^{\mu} &= M^{\mu}_{\phantom{\mu}\nu}x^{\nu},\nonumber\\
  A_{\mu}(x) &= \phi_{\nu}(y)M^{\nu}_{\phantom{\nu}\mu},
\label{eq:M-transf}
\end{align}
and in the Euclidean case
\begin{align}
  y^{\mu} &= M_{E\mu\nu}\Pi_{\nu\rho}x_{E\rho},\nonumber\\
  A_{E\mu}(x_E) &= \phi_{\rho}(y)M_{E\rho\nu}\Pi_{\nu\mu},
\label{eq:E-transf}
\end{align}
where $M$ and $M_E$ are the diagonal matrices
\begin{align}
\label{eq:matrix}
  M^{\mu}_{\phantom{\mu}\nu} &=
  \diag(\frac{T_0}{T}\frac{1}{\sqrt{2}\cosh(\chi/2)},
  \frac{T_0}{T}\frac{1}{\sqrt{2}\sinh(\chi/2)},1,1),\nonumber\\
  M_{E\mu\nu} &=
  \diag(\frac{T_0}{T}\frac{1}{\sqrt{2}\cos(\theta/2)},
  \frac{T_0}{T}\frac{1}{\sqrt{2}\sin(\theta/2)},1,1),
\end{align}
and $\Pi$ simply permutes the Euclidean coordinates to put
them in the order $0123$,
\begin{equation*}
  \Pi_{\mu\nu} = \left(
  \begin{array}{cccc}
    0 & 0 & 0 & 1\\
    1 & 0 & 0 & 0\\
    0 & 1 & 0 & 0\\
    0 & 0 & 1 & 0
  \end{array}
  \right).
\end{equation*}
The Wilson loops are then changed into
\begin{align}
{\cal W}^{(T)}_{1,2}[A] &=  
{\dfrac{1}{N_c}} \tr \left\{ {\cal P} \exp
\left[ -ig \displaystyle\oint_{{\Gamma}_{1,2}} \phi_{\mu}(y) dy^{\mu} \right]
\right\} \equiv W_{\Gamma_{1,2}}[\phi] ,
\nonumber\\
\widetilde{\cal W}^{(T)}_{1,2}[A_E] &= 
{\dfrac{1}{N_c}} \tr \left\{ {\cal P} \exp
\left[ -ig \displaystyle\oint_{{\Gamma}_{1,2}} \phi_{\mu}(y) dy^{\mu} \right]
\right\} \equiv W_{\Gamma_{1,2}}[\phi],
\end{align}
where the new paths are
\begin{align}
{\Gamma}_1 :& \quad
Y_{1q}^\mu(\tau) = z^\mu +
\frac{\delta^{\mu}_{\phantom{\mu}0}+\delta^{\mu}_{\phantom{\mu}1}}{\sqrt{2}}
\tau + (1-f_1) R_1^\mu, ~~~~  \nonumber \\
\phantom{{\Gamma}_1 :} & \quad
Y_{1\bar{q}}^\mu(\tau) = z^\mu +
\frac{\delta^{\mu}_{\phantom{\mu}0}+\delta^{\mu}_{\phantom{\mu}1}}{\sqrt{2}}
\tau - f_1 R_1^\mu , 
\nonumber \\
{\Gamma}_2 :& \quad
Y_{2q}^\mu(\tau) =
\frac{\delta^{\mu}_{\phantom{\mu}0}-\delta^{\mu}_{\phantom{\mu}1}}{\sqrt{2}}
\tau + (1-f_2) R_2^\mu, ~~~~  \nonumber \\
\phantom{{\Gamma}_2 :}& \quad
Y_{2\bar{q}}^\mu(\tau) =
\frac{\delta^{\mu}_{\phantom{\mu}0}-\delta^{\mu}_{\phantom{\mu}1}}{\sqrt{2}}
\tau - f_2 R_2^\mu , 
\end{align}
with $\tau\in [-T_0,T_0]$, and closed by the usual
transverse straight--line paths at $\tau= \pm T_0$. 

We have written the loops in the two cases as the same functional of
the new variables, but the transformations that we performed are
different, giving rise to different actions; we make this explicit
by introducing the notation 
\begin{equation}
  \label{eq:fintegral}
  \la {\cal O}[\phi] \ra_S \equiv \dfrac{\displaystyle\int[D\phi]{\cal O}[\phi]
  e^{-S[\phi]}}{\displaystyle\int[D\phi]e^{-S[\phi]}},
\end{equation}
and writing for the correlation functions and expectation values in
the two theories
\begin{align}
  &\la {\cal W}^{(T)}_{1}{\cal W}^{(T)}_{2} \ra = \la W_{\Gamma_1}
  W_{\Gamma_2} \ra_{-iS_M^{\rm Y.M.}},   &
  &\la {\cal W}^{(T)}_{i} \ra = \la W_{\Gamma_i} \ra_{-iS_M^{\rm Y.M.}}, \nonumber\\ 
  &\la \widetilde{\cal W}^{(T)}_{1}\widetilde{\cal W}^{(T)}_{2} \ra_E =
  \la W_{\Gamma_1} W_{\Gamma_2} \ra_{S_E^{\rm Y.M.}}, &
  &\la \widetilde{\cal W}^{(T)}_{i} \ra_E = \la W_{\Gamma_i} \ra_{S_E^{\rm Y.M.}} ,
  \label{eq:trans_corrfunc}
\end{align}
where $S_M^{\rm Y.M.}$ and $S_E^{\rm Y.M.}$ are the transformed Minkowskian and Euclidean
pure--gauge (Yang--Mills) actions:
\begin{align}
  \label{eq:mod_M_ac}
  S_M^{\rm Y.M.} &= -\frac{1}{2}\sum_{\mu,\nu=0}^3  C_{M\mu\nu}(\chi,T)\int d^4y
  \tr(\Phi_{\mu\nu})^2,\\ 
  \label{eq:mod_E_ac}
  S_E^{\rm Y.M.} &= \frac{1}{2}\sum_{\mu,\nu=0}^3 {C}_{E\mu\nu}(\theta,T)\int d^4y
  \tr(\Phi_{\mu\nu})^2.
\end{align}
Here $(\Phi_{\mu\nu})^2$ is understood as the square of the hermitian
matrix $\Phi_{\mu\nu}$,
\begin{equation}
  \Phi_{\mu\nu} = \dde{\mu}\phi_{\nu} - \dde{\nu}\phi_{\mu} +
  ig[\phi_{\mu},\phi_{\nu}],
\end{equation}
and the symmetric coefficients $C_{M\mu\nu}$ and ${C}_{E\mu\nu}$
are 
\begin{align}
  &\left\{\begin{aligned}
  C_{M01} & = -C_{M23}^{-1} = -\left(\dfrac{T_0}{T}\right)^2\dfrac{1}{|\sinh\chi|}, \\
  C_{M02} & = C_{M03} = -C_{M12}^{-1} = -C_{M13}^{-1}=
  -\dfrac{|\sinh\chi|}{\cosh\chi+1},
  \end{aligned}\right. \label{coeff_1}\\
  \intertext{and}
  &\left\{\begin{aligned}
  {C}_{E01} & = {C}_{E23}^{-1}  =
  \left(\dfrac{T_0}{T}\right)^2\dfrac{1}{|\sin\theta|}, \\
  {C}_{E02} & = {C}_{E03} =  {C}_{E12}^{-1}
  = {C}_{E13}^{-1} =  \dfrac{|\sin\theta|}{\cos\theta+1},
  \end{aligned}\right.\label{coeff_2}
\end{align}
and $C_{M\mu\mu}={C}_{E\mu\mu}=0$ $\forall \mu$. 

If we now restrict the angular variables to the intervals
$\chi\in {\cal I}_M = \mathbb{R}^+$ and $\theta\in {\cal
  I}_E=(0,\pi)$ (see Ref.~\cite{crossing}), we can drop the absolute
values in Eqs.~(\ref{coeff_1}) and (\ref{coeff_2}), obtaining
coefficient functions which can be analytically  
extended throughout the respective complex planes in both
variables, with the possible exception (depending on the specific
coefficient) of the isolated singular points (poles)
$T=0,\infty$, and $\chi=0,\infty$ in the Minkowskian case or
$\theta=n\pi$, $n\in\mathbb{Z}$ in the Euclidean case.  To avoid
confusion, 
we will denote with an overbar the {\it analytic extensions}
$\overline{C}_{M\mu\nu}$ and $\overline{C}_{E\mu\nu}$ obtained
starting from ${\cal I}_M$ and ${\cal I}_E$ at real positive $T$, in
the two cases respectively\footnote{By construction, the two quantities
  $C_{M,E}$ and $\overline{C}_{M,E}$ (and thus also the correlation
  functions and their analytic extensions) coincide in ${\cal
    I}_{M,E}$ at positive real $T$; nevertheless, as already pointed
  out in~\cite{crossing}, one easily sees that $C_{M\mu\nu} \neq
  \overline{C}_{M\mu\nu}$ for negative values of $\chi$, and similarly
  for the Euclidean coefficients $C_{E\mu\nu} \neq
  \overline{C}_{E\mu\nu}$ for $\theta\in (\pi,2\pi)$. }.  

In the next section we will discuss the relevant analyticity issues on
the basis of the functional--integral representation just obtained.

\section{Analytic continuation}

The functional integral is, as it stands, a mathematically
ill--defined object, and it acquires a precise meaning only through
the specification of a practical prescription to calculate
it. In perturbation theory, for example, the functional integral is
essentially a book--keeping device for the perturbative
expansion, and formal manipulations of the integral correspond to
well--defined operations on each term of the series. However, such a
series is known not to be convergent, and moreover it can be obtained
without any reference to the functional integral: to give the latter 
an intrinsic meaning, one should go beyond perturbation theory. 

In the most common nonperturbative approach, the lattice approach, one
replaces the infinite space--time continuum with a lattice of finite
volume $V$ and finite spacing $a$: the starting point is thus an
ordinary multidimensional integral, for which the ordinary theorems of
calculus apply, and the functional integral is defined as the
$V\to\infty$, $a\to 0$ 
limit of this quantity. For gauge theories one can exploit
the freedom in the choice of the lattice action in order to preserve
the gauge symmetry at any stage of the calculation~\cite{Wil}: this is
the case we will have in mind whenever referring to the lattice
approach in the following.

In this section we will discuss the analyticity properties of the
correlation function ${\cal G}_E$, using a formal argument based on
the functional integral representation of the previous section; the validity
of the argument will be justified in subsection 4.4 using a lattice 
regularisation.

\subsection{Analyticity domain of the Euclidean correlation function}

A function of a complex variable is analytic if its derivative exists
in complex sense. As the correlation function ${\cal G}_E$
is known in terms of a functional integral, the question is under
which conditions we are allowed to bring the derivative under the sign
of integral, for in that case we can infer the analytic properties of
the correlation function directly from its functional--integral
representation. 
In the case of ordinary integrals 
one can bring the derivative with respect to a parameter under the
integral sign as long as the resulting integral is
convergent\footnote{To be precise, pointwise convergence is generally
  not sufficient, while uniform convergence is a sufficient
  condition.}; in the case of functional integrals we assume that
this remains true. The following argument is then formal, 
but it can be made more rigorous treating the functional integral
in some regularisation scheme\footnote{A {\it mathematically
  rigorous} proof of analyticity should also show that the removal of
  the regularisation does not spoil the results; however, the solution of this
  problem is currently out of reach.}; this will be done in subsection 4.4.

The functional integrals defined by means of
Eq.~(\ref{eq:fintegral}) are expected to be convergent as long as
the real part of the action is positive--definite, for in this case
the exponential factor strongly suppresses configurations with large
action. We have already seen that the transformed Euclidean action
$S_E^{\rm Y.M.}$ of Eq.~(\ref{eq:mod_E_ac}), as a function of $\theta$ and $T$,
is analytic in the whole two--dimensional $(\theta,T)$ complex space, 
with the exception of some isolated singular points; if we now set
$S=S_E^{\rm Y.M.}$ in Eq.~(\ref{eq:fintegral}) and allow derivatives to pass under
the (functional) integral sign, for operators ${\cal O}$ which do not
depend on $\theta$ and $T$ we have
\begin{multline}
  \frac{\de}{\de\theta} \la {\cal O}[\phi] \ra_{S_E^{\rm Y.M.}} = \\
  \la -\frac{\de S_E^{\rm Y.M.}}{\de\theta}{\cal O}[\phi] \ra_{S_E^{\rm Y.M.}} -  \la
  -\frac{\de S_E^{\rm Y.M.}}{\de\theta} \ra_{S_E^{\rm Y.M.}} \la{\cal O}[\phi] \ra_{S_E^{\rm Y.M.}}
\end{multline}
(and similarly for the derivative with respect to $T$), where
$\frac{\de S_E^{\rm Y.M.}}{\de\theta}$ is the space--time integral of
a polynomial in the fields and their derivatives, with coefficients analytic in
$\theta$ and $T$, which does not change the convergence properties of the 
functional integral. We conclude (formally) that the correlation
function ${\cal G}_E$ can be analytically extended to complex values
of $\theta$ and $T$ for which the real part of the action $S_E^{\rm
  Y.M.}$ is positive--definite, and this happens if and only if the
{\it convergence conditions}  
\begin{equation}
  \label{eq:convcond}
  \Re\overline{C}_{E\mu\nu}(\theta,T) > 0 \quad \forall \mu,\nu
\end{equation}
for the (analytically extended) coefficients are satisfied.

Singular points of the coefficients are artifacts of our functional
integral representation, and they are not necessarily singular
points of the correlation function: indeed, while singularities are
expected at the points $\theta=0$ and
$\theta=\pi$ on the basis of the relation between the
correlation function ${\cal G}_E$ and the static dipole--dipole
potential~\cite{Pot} (see also~\cite{crossing}), no singularity is
expected at $T=0$, where ${\cal G}_E$ is expected to vanish. These
points will be excluded from the following analysis, with the
exception of $T=\infty$ which will be considered separately\footnote{A
  true singularity could appear if the Wilson--loop expectation value
  vanishes for some choice of complex $\theta$ and $T$: in the
  following we will not discuss this possibility, although we cannot
  rule it out.}.

We solve now the {\it convergence conditions} (\ref{eq:convcond}),
substituting $\theta$ with the complex variable $z\equiv\theta-i\chi$
(with real $\theta$ and $\chi$) and writing for the complex variable
$T$, $T=|T|e^{i\psi/2}$; as $\sign(\Re 1/z) = \sign(\Re z)$, it 
suffices to study the inequalities
\begin{align}
&\left\{
\begin{aligned}
  &\Re \left[e^{i\psi}\sin(\theta-i\chi)\right] > 0,\\
  &\Re \left[\dfrac{\sin(\theta-i\chi)}{1+\cos(\theta-i\chi)}\right] > 0,
\end{aligned}\right.
\intertext{which are equivalent to}
&\left\{
\begin{aligned}
  & F(\theta,\chi,\psi) \equiv e^{\chi}\sin(\theta+\psi) +
  e^{-\chi}\sin(\theta-\psi) > 0, \\
  & \sin\theta(\cosh\chi + \cos\theta) > 0.
\end{aligned}\right.
\end{align}
Since $F(\theta,\chi,\psi+2\pi)=F(\theta,\chi,\psi)$, it suffices to
consider $-\pi\le\psi\le\pi$ only.
The second inequality is satisfied only for $\theta\in (0,\pi)$
\footnote{Actually it is satisfied for  $\theta\in (2k\pi,2k\pi+\pi)$,
  but we are interested in a connected analyticity domain.}, thus
obstructing the analytic extension outside the region
$\{\theta\in(0,\pi), \chi\in \mathbb{R}, T\in\mathbb{C}\}$. The first
inequality further restricts this domain: noting the relations
\begin{multline}
F(\theta,\chi,\psi) = \\ F(\theta,-\chi,-\psi) = F(\pi-\theta,\chi,-\psi)
= -F(\pi-\theta,\chi,\pi-\psi)
\label{eq:dom-ident}
\end{multline}
we infer that the domain must be symmetric under the transformations
$(\chi,\psi)\to (-\chi,-\psi)$ and $(\theta,\psi)\to (\pi-\theta,-\psi)$,
and that we can study the inequality for $\theta\in(0,\pi/2], \chi>0$ only
and then extend the results. In this region the inequality is satisfied
for
\begin{equation}
  \{ \psi \ge \pi/2, \quad \theta <
  \atan\left({\tan(\pi-\psi)}{\tanh \chi}\right)\} \cup \{\psi
  < \pi/2 \},
\end{equation}
which implies for $\theta\in[\pi/2,\pi), \chi>0$ the condition
\begin{equation}
  \{\psi < \pi/2, \quad \theta < \pi - \atan\left({\tan\psi}{\tanh
  \chi}\right) \}.
\end{equation}
Summarising, if we define
\begin{equation}
     B(\chi,\psi)\equiv \atan\left({\tan(\pi-\psi)}{\tanh \chi}\right) +
     \pi\Theta\left(\frac{\pi}{2}-\psi\right),
     \label{eq:B}
\end{equation}
where $\Theta(x)$ is the Heaviside step function,
for fixed positive $\chi$ and positive $\psi$ one has to satisfy  
\begin{equation}
  0 < \theta < B(\chi,\psi),
\end{equation}
while for fixed positive $\chi$ and negative $\psi$ one has
\begin{equation}
   \pi - B(\chi,-\psi) < \theta < \pi;
\end{equation}
finally, for negative $\chi$
\begin{align}
  & \pi - B(-\chi,\psi) < \theta < \pi, \quad \psi>0,\nonumber\\
  &  0 < \theta < B(-\chi,-\psi), \quad \psi<0.
\end{align}
At $\chi=0$ the allowed range is $|\psi|<\pi/2, \theta\in(0,\pi)$.
The boundary of the domain at large $|\chi|$ is easily found noting that
\begin{equation}
   \lim_{\chi\to +\infty}B(\chi,\psi) = \pi-\psi.
\end{equation}

The previous inequalities define a connected subset ${\cal V}$ of the
3D real $(\theta,\chi,\psi)$--space; moreover, as the modulus
$|T|$ never enters in the previous equations, the section of the
analyticity domain is the same irrespectively of $|T|$. No dependence on
the arbitrary parameter $T_0$ is found, too, as expected. We have
thus found a connected analyticity domain ${\cal D}_E$,
\begin{equation}
  \label{eq:domain}
  {\cal D}_E  = \{(z,T)\in \mathbb{C}^2 \,|\, (\theta,\chi,\psi)\in {\cal V}\}
\end{equation}
for the extension of the Euclidean correlation function from $\theta\in{\cal
  I}_E$ at positive real $T$.

\begin{figure}[h]
  \centering
  \includegraphics[width=1\textwidth]{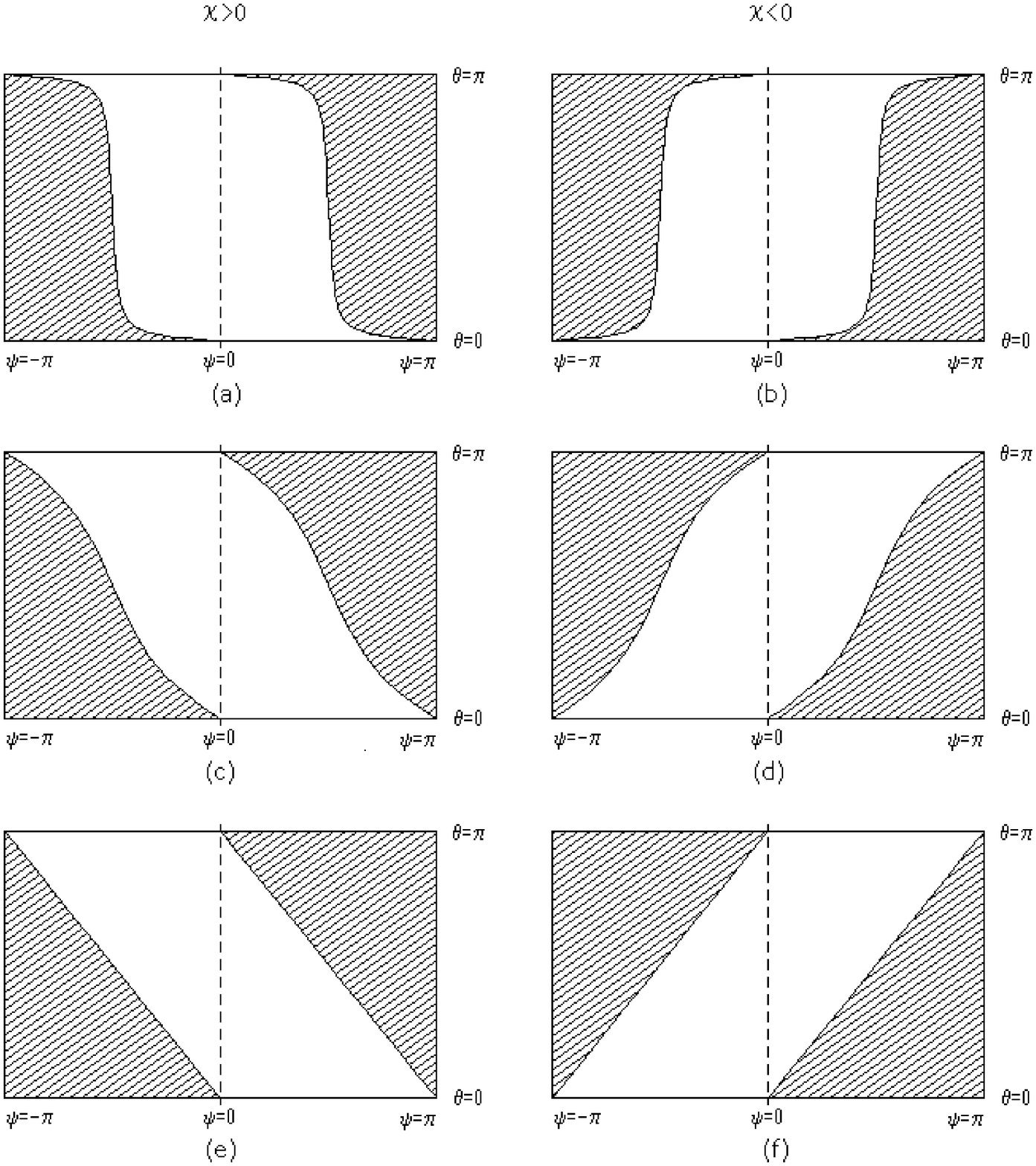}
  \caption{Section of the analyticity domain of ${\cal G}_E$ at fixed
    $\chi$ (white area) for various values of $\chi$: (a) $\chi=0.06$;
    (b) $\chi=-0.06$; (c) $\chi=0.6$; (d) $\chi=-0.6$;
    (e) $\chi\to +\infty$; (f) $\chi\to -\infty$.} 
  \label{fig:fixchi}
\end{figure}

Sections of this subset at fixed $\chi$ are shown in
Fig.~\ref{fig:fixchi}. The domain ``thins out'' as one tends towards
$\psi\to\pi$ or $\psi\to -\pi$; according to previous
works~\cite{Meggiolaro97,Meggiolaro98,Meggiolaro02,Meggiolaro05,crossing,Meggiolaro07}, 
we expect to find the ``direct'' physical region [i.e.,  
  the Minkowskian action of Eq.~(\ref{eq:mod_M_ac})] at
$\theta=0+,\chi>0,\psi=\pi$, and the ``crossed'' physical region at
$\theta=\pi-,\chi<0,\psi=\pi$. This issue 
will be investigated in the next subsection, where we discuss also
what is found at the other edges of the analyticity domain. 

\begin{figure}[h]
  \centering
  \includegraphics[width=\textwidth]{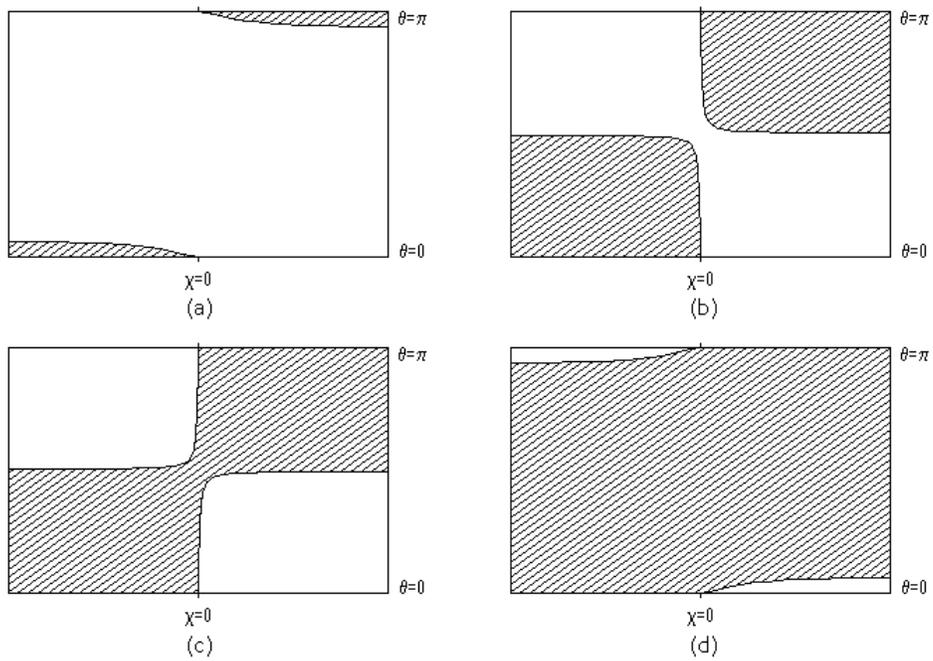}
  \caption{Section of the analyticity domain of ${\cal G}_E$ in the
    $\chi$--$\theta$ plane (white area), for various values of $\psi$:
  (a) $\psi=0.2$; (b) $\psi=\pi/2-0.02$; (c)  $\psi=\pi/2+0.02$; (d)
  $\psi=\pi-0.2$.}  
  \label{fig:fixpsi}
\end{figure}

For clarity reasons, and also to make contact with the analysis
performed in previous
works~\cite{Meggiolaro97,Meggiolaro98,Meggiolaro02,Meggiolaro05,crossing,Meggiolaro07},  
sections of the same analyticity domain at fixed $\psi$ are
shown in Fig.~\ref{fig:fixpsi}: the whole
``strip'' ${\cal S}_E\equiv\{z=\theta-i\chi\, |\, \theta\in
(0,\pi),\chi\in\mathbb{R}\}$ (at $\psi=0$) reduces to disjoint regions
near the edges of the domain (at $\psi\simeq\pm\pi$).

\subsection{Analytic continuation, crossing symmetry and the
  ``reflection relation''}

It is convenient to denote the two ``physical'' edges of the
analyticity domain as $E^{\rm dir}$ and $E^{\rm cross}$, with
($z=\theta-i\chi$, $T=|T|e^{i\psi/2}$)
\begin{align}
E^{\rm dir} &=  \{(z,T)\in \mathbb{C}^2\, |\,
\theta=0,\chi\in\mathbb{R}^+,\psi=\pi\},\nonumber\\
E^{\rm cross} &=
\{(z,T)\in \mathbb{C}^2 \,|\, \theta=\pi,\chi\in\mathbb{R}^-,\psi=\pi\};  
\end{align}
it is also convenient to adopt the notation $E^* = \{(z,T) \,|\,
(z^*,T^*)\in E\}$,  
with which the other two edges of the domain are denoted as $E^{\rm
  dir}{}^*$ and $E^{\rm cross}{}^*$.

As we let $\psi\to\pi$ and $\theta\to 0$ (from positive values),
approaching $E^{\rm dir}$ from the inside, 
the coefficients $\overline{C}_{E\mu\nu}$ become imaginary,
and 
\begin{equation}
  \overline{C}_{E\mu\nu}(-i\chi,iT) =
  i{C}_{M\mu\nu}(\chi,T) 
\end{equation}
so that
\begin{equation}
  S_E^{\rm Y.M.} \stackrel{\theta\to -i\chi,\,T\to iT}{\longrightarrow} -iS_M^{\rm Y.M.},
\end{equation}
i.e., according to Eq.~(\ref{eq:trans_corrfunc}),
\begin{equation}
  {\cal G}_M (\chi;T;\vec{z}_\perp;1,2)=
  \overline{\cal G}_E(-i\chi;iT;\vec{z}_\perp;1,2), 
  \qquad \forall \chi\in\mathbb{R}^+,T\in\mathbb{R}^+.
\label{eq:an_cont_phys}
\end{equation}
We thus find that Minkowskian and Euclidean correlation functions are
connected by the expected analytic
continuation~\cite{Meggiolaro97,Meggiolaro98,Meggiolaro02,Meggiolaro05},
of which we have given here an alternative derivation. More precisely, we can
define the analytic extension of the Minkowskian correlation function ${\cal
  G}_M$ from $\chi\in\mathbb{R}^+,T\in\mathbb{R}^+$ to $w\equiv\chi +
i\theta=iz$ and complex $T$, by setting
\begin{equation}
    \overline{\cal G}_M(w;T;\vec{z}_\perp;1,2)
\equiv \overline{\cal G}_E (-iw;iT;\vec{z}_\perp;1,2) 
\qquad \forall (w,T)\in {\cal D}_M,
\label{eq:an_cont_ext_M}
\end{equation}
where ${\cal D}_M = \{(w,T)\in\mathbb{C}^2\,|\, (-iw,iT)\in{\cal D}_E\}$,
and ${\cal D}_E$ has been determined in the previous section, see
Eq.~(\ref{eq:domain}). The right--hand side of
Eq.~(\ref{eq:an_cont_ext_M}) is indeed the analytic extension of
${\cal G}_M$, as the two functions coincide at real positive values of
$w$ and real positive $T$, by virtue of Eq.~(\ref{eq:an_cont_phys}).
Alternatively, the analytic--continuation relation
Eq.~(\ref{eq:an_cont_ext_M}) for the extended functions can be written
as  
\begin{equation}
  \overline{\cal G}_E(z;T;\vec{z}_\perp;1,2)
= \overline{\cal G}_M (iz;-iT;\vec{z}_\perp;1,2) ,
\qquad \forall (z,T)\in {\cal D}_E.
\label{eq:an_cont_ext_E}
%\label{analytic}
\end{equation}
From the point of view of the Minkowskian analytically--extended correlation
function $\overline{\cal G}_M$, the physical axis for the complex angular
variable $w=\chi+i\theta$ is approached from above, i.e., from positive
imaginary values; as $\chi\simeq \log (s/m^2)$ at high energies, this
corresponds to $\chi + i\epsilon = \log (s/m^2) + i\epsilon =  \log
[s/(m-i\epsilon)^2]$, in agreement with the usual ``$-i\epsilon$''
prescription~\cite{Meggiolaro07}.

According to the {\it crossing--symmetry relations}
(\ref{eq:crossrel}) (derived in~\cite{crossing}), we should find the
physical amplitude in the ``crossed'' channel at negative values of
$\chi$ as $\psi\to\pi$ and $\theta\to \pi$, i.e., at the edge  $E^{\rm
  cross}$ of the analyticity domain. Here we find 
\begin{equation}
  \overline{C}_{E\mu\nu}(\pi-i\chi, iT) =
  \sum_{\alpha,\beta=0}^3 iS_{\mu\alpha}S_{\nu\beta}{C}_{M\alpha\beta}(-\chi,T),
\end{equation}
where the matrix $S$ simply interchanges the $0$ and $1$
components of fields and coordinates,
\begin{equation}
  S_{\mu\nu} = \left(
  \begin{array}{cccc}
    0 & 1 & 0 & 0\\
    1 & 0 & 0 & 0\\
    0 & 0 & 1 & 0\\
    0 & 0 & 0 & 1
  \end{array}\right).
\end{equation}
We can reabsorb the matrix $S$ into the loops with a transformation of
fields and coordinates,
with the only effect of reversing the orientation of $W_{\Gamma_2}$,
so that $W_{\Gamma_2}\to W_{\Gamma_2}^*$, all the rest remaining
unchanged; we thus find that the Euclidean correlation function 
is analytically continued to the physical correlation function (with 
positive hyperbolic angle $-\chi$) of a loop and an antiloop, as
expected~\cite{crossing}. As a by--product, we reobtain the
crossing--symmetry relation for the loops (\ref{eq:crossrel}), which
can now be extended to the whole analyticity domain\footnote{To make the
  statement of Ref.~\cite{crossing} more precise, 
we notice that, although the {\it same} complex $T$ appears on both
sides of Eq.~(\ref{eq:cross_ext}), these relations cannot be obtained
by a {\it simple} analytic continuation $z\to \pi -z$ (or $w\to i\pi -w$) in
the angular variable only at {\it fixed} $T$. Indeed, Fig.~\ref{fig:fixpsi}(d)
shows that the section of the analyticity domain at constant $\psi\simeq\pi$
is made up of two disconnected regions near $E^{\rm dir}$ and  $E^{\rm
  cross}$, so that a {\it double} analytic continuation, both in the angular
variable and in $T$, is needed to prove Eq.~(\ref{eq:cross_ext}).}, 
\begin{eqnarray}
  \overline{\mathcal{G}}_M(i\pi-w;T;\vec{z}_{\perp};1,2)
&=&\overline{\mathcal{G}}_M(w;T;\vec{z}_{\perp};1,\overline{2})
\nonumber \\ 
&=&\overline{\mathcal{G}}_M(w;T;\vec{z}_{\perp};\overline{1},2) ,
\quad \forall (w,T) \in {\cal D}_M,
\nonumber \\
  \overline{\mathcal{G}}_E(\pi-z;T;\vec{z}_{\perp};1,2)
&=&\overline{\mathcal{G}}_E(z;T;\vec{z}_{\perp};1,\overline{2})
\nonumber\\  
&=&\overline{\mathcal{G}}_E(z;T;\vec{z}_{\perp};\overline{1},2) ,
\quad\forall (z,T)\in {\cal D}_E,
\label{eq:cross_ext}
\end{eqnarray}
since ${\cal D}_E$ satisfies $(z,T)\in {\cal D}_E \Leftrightarrow
(\pi-z,T)\in {\cal D}_E$ [this is easily seen by combining the first
  two relations of Eq.~(\ref{eq:dom-ident}) into
  $F(\theta,\chi,\psi)=F(\pi-\theta,-\chi,\psi)$], and thus $(w,T)\in
{\cal D}_M \Leftrightarrow (i\pi-w,T)\in {\cal D}_M$. 

To see what happens at the other two edges of the analyticity domain
it suffices to notice that the domain ${\cal D}_E$ possesses the
symmetry ${\cal D}_E={\cal D}_E^*$ [see the first relation in
Eq.~(\ref{eq:dom-ident})], and that the coefficients 
$\overline{C}_{E\mu\nu}$ satisfy the {\it reflection relation} 
\begin{equation}
  \label{eq:refl0}
  \overline{C}_{E\mu\nu}(z^*,T^*) = \overline{C}_{E\mu\nu}(z,T)^*.
\end{equation}
We thus find that the correlation function takes conjugate values at
conjugate points $(z,T)$ and $(z^*,T^*)$, as $C$--invariance (which is
not lost when we perform the field transformations) implies
that a correlation function does not change if we substitute all the
Wilson loops with their antiloops (the notation should be clear):
\begin{multline}
  \la W_{\Gamma_1} W_{\Gamma_2} \ra_{S_E^{\rm Y.M.}[\overline{C}_{E}(z^*,T^*)]} =
  \la W_{\Gamma_1} W_{\Gamma_2} \ra_{S_E^{\rm Y.M.}[\overline{C}_{E}(z,T)^*]} =\\
  \la W_{\Gamma_1}^* W_{\Gamma_2}^* \ra_{S_E^{\rm Y.M.}[\overline{C}_{E}(z,T)^*]}
  = \la W_{\Gamma_1} W_{\Gamma_2} \ra_{S_E^{\rm Y.M.}[\overline{C}_{E}(z,T)]}^*
\end{multline}
(and similarly for the loop expectation values). We thus conclude that
also the correlation function satisfies the {\it reflection relation}
\begin{equation}
  \label{eq:reflG}
   \overline{\cal G}_E(z^*;T^*;\vec{z}_\perp;1,2) =  \overline{\cal
     G}_E(z;T;\vec{z}_\perp;1,2)^*.
\end{equation}
In particular, this means that at \mbox{$\psi=-\pi$} we find the
complex conjugate of the physical correlation functions, respectively
at  $E^{\rm dir}{}^*$ ($\chi<0$) for the ``direct channel'' and at
$E^{\rm cross}{}^*$ ($\chi>0$) for the ``crossed channel''. Moreover,
from the previous  relation we find that the Euclidean correlation
function at $\chi=0,\psi=0$ is a real function, as can be shown also
in a more direct way making use of the $C$--invariance of the usual
Yang--Mills action (this has already been noticed in~\cite{lattice}).

\subsection{Analyticity properties of the correlation function with the
IR cutoff removed}

As the physically relevant quantities are the correlation functions
with the IR cutoff removed~\cite{BL,Meggiolaro05},
\begin{align}
 {\cal C}_M(\chi;\vec{z}_\perp;1,2) &\equiv \lim_{T\to\infty} {\cal
  G}_M(\chi;T;\vec{z}_\perp;1,2),\nonumber \\
{\cal C}_E(\theta;\vec{z}_\perp;1,2) & \equiv \lim_{T\to\infty}{\cal
  G}_E(\theta;T;\vec{z}_\perp;1,2),
\end{align}
we will discuss now what can be inferred about their analyticity
properties from the properties of ${\cal G}_E$.

The results of the previous section (see Fig.~\ref{fig:fixchi}) show
that $\overline{\cal G}_E$, as a function of the complex variable $T$ at fixed
$z= \theta-i\chi$, is analytic in the sector $-\pi/2 + \Delta < \arg T=\psi/2
< \Delta$, where $\Delta=\Delta(z)\in (0,\pi/2)$. The precise form of $\Delta$ 
is not needed here, but it can be obtained solving for $\Delta$ the
equation $\theta=B(\chi,2\Delta)$, with $B$ defined in Eq.~(\ref{eq:B}).
Note that the sector extends on an 
angle $\pi/2$, irrespectively of $\Delta$ (i.e., of $\theta$ and
$\chi$), and that the ``strip'' $\psi=0$  falls completely inside
the domain, so that one can define $I_\Delta\equiv (-\pi + 2\Delta,2\Delta)$,
and rewrite ${\cal D}_E$ as ${\cal D}_E = \{(z,T)\,|\,z\in {\cal S}_E
,\,\psi\in I_{\Delta(z)}\}$; in the same notation we have ${\cal
  D}_M = \{(w,T)\,|\,w\in {\cal S}_M ,\,\psi\in I_{\Delta(-iw)}-\pi\}$,
where ${\cal S}_M \equiv i{\cal S}_E = \{w=\chi+i\theta\, |\,
\chi\in\mathbb{R},\theta\in (0,\pi)\}$ is the Minkowskian ``strip''.

A simple nonperturbative argument for the IR finiteness of the
normalised correlation function in a non--Abelian gauge theory is as
follows\footnote{In the 
  case of Abelian pure--gauge theory (i.e., {\it quenched} QED) the
  limit has been shown to be finite by direct computation
  in~\cite{Meggiolaro05}, and the usual analytic continuation
  (\ref{analytic_C}) in the angular variable only is explicitly seen
  to be the correct one.}. 
Due to the short--range nature of strong interactions, those parts of
the partons' trajectories that lie too far aside with respect to the
``vacuum correlation length'' (see Ref.~\cite{DDSS} and references
therein) 
do not affect each other; translated in terms of the
functional--integral description of the process, this means that there
should be a  ``critical'' length $T_c$,
beyond which the normalised correlation function becomes independent
of $T$. Indeed, the available lattice data confirm that ${\cal G}_E$
becomes approximately constant for large (real positive) values of
$T$~\cite{lattice}. As the existence of a ``vacuum correlation
length'' is usually ascribed to the non--trivial dynamics dictated by
non--Abelian gauge invariance, the previous argument is expected to
apply also for the analytically--extended correlation functions,
substituting the real variable $T$ with the modulus of the complex
variable $|T|$. 

In conclusion, the analytically extended correlation
function is expected to be analytic and, by the above--mentioned
argument, also bounded (at least for large enough $|T|$), as a function of the
complex variable $T$, in a sector of the corresponding complex plane, 
enclosed between two straight lines departing from the origin at an
angle $\pi/2$, with finite limits as $|T|\to\infty$ along the two
straight lines. We can then apply the Phragm\'en--Lindel\"of
theorem (see theorem 5.64 of Ref.~\cite{Tit}) to show that
$\overline{\cal G}_E$ converges uniformly to a unique 
value in the whole sector as $|T|\to\infty$, and define
unambiguously the functions 
\begin{align}
  \overline{\cal C}_M(w;\vec{z}_\perp;1,2) &\equiv \lim_{|T|\to\infty}
  \overline{\cal G}_M(w;T;\vec{z}_\perp;1,2),\qquad \forall w\in {\cal S}_M \nonumber \\
  \overline{\cal C}_E(z;\vec{z}_\perp;1,2) & \equiv \lim_{|T|\to\infty}
  \overline{\cal G}_E(z;T;\vec{z}_\perp;1,2), \qquad \forall z\in {\cal S}_E,
\end{align}
since the limit on the right--hand side does not depend on the
particular direction in 
which one performs it. One easily sees that $\overline{\cal C}_M$ and
$\overline{\cal C}_E$ are the analytic extensions of ${\cal C}_M$ and of
${\cal C}_E$. Indeed, in the Minkowkian case, it suffices to take the limit
$|T|\to\infty$ in the equations above setting $w=\chi\in\mathbb{R}^+$, as in this
case the sector in the complex--$T$ plane, for which $\overline{\cal G}_M$ is
analytic, extends up to real positive values of $T$. In the Euclidean case,
since real positive $T$ are always inside the domain ${\cal D}_E$ for every
value of the {\it complex} angular variable $z=\theta-i\chi$ in the ``strip'' 
${\cal S}_E$, one can think of the limit $|T|\to\infty$ as being made
along the real axis in the positive direction for every $z$, and
it thus suffices to take $z$ to be real, $z=\theta$, and in the
interval ${\cal I}_E$. If we now take the limit $|T|\to\infty$ in the
analytic continuation relations, Eqs.~(\ref{eq:an_cont_ext_M}) and
(\ref{eq:an_cont_ext_E}), we obtain the analytic continuation
relations with the IR cutoff removed~\cite{Meggiolaro05}, 
\begin{eqnarray}
  \overline{\cal C}_M(w;\vec{z}_\perp;1,2) &=&
  \overline{\cal C}_E(-iw;\vec{z}_\perp;1,2) ,
  \qquad \forall w\in {\cal S}_M,
  \nonumber \\
  \overline{\cal C}_E(z;\vec{z}_\perp;1,2) &=&
  \overline{\cal C}_M(iz;\vec{z}_\perp;1,2) ,
  \qquad \phantom{-}\forall z\in {\cal S}_E.
\end{eqnarray}
The crossing--symmetry relations are still valid for $\overline{\cal
  C}_M$ and $\overline{\cal C}_E$ throughout the respective
analyticity domains ${\cal S}_M$ and ${\cal S}_E$, as one can prove by
taking the limit  $|T|\to\infty$ in Eq.~(\ref{eq:cross_ext}) (relying
again on the Phragm\'en--Lindel\"of theorem mentioned above). Note
also that ${\cal C}_E(z^*)={\cal  C}_E(z)^*$ throughout the domain
of analyticity, as one can easily see by taking $|T|\to\infty$ in
Eq.~(\ref{eq:reflG}). This conclusion can 
be reached independently, showing that ${\cal C}_E(z)$ is real
for real $z$ (as briefly explained at the end of the previous
subsection), and using Schwartz's reflection principle, but in this way
no insight on the analyticity domain is obtained.

\subsection{Lattice regularisation}

As already pointed out, the functional integral must be regularised to
become a well--defined mathematical object; here we justify the
formal argument given above using a lattice regularisation. 
In this approach the ill--defined continuum functional integral is
replaced with a well--defined (multidimensional) integral, which in
the case of gauge theories can be chosen to be an integral on the
gauge group manifold~\cite{Wil}, 
\begin{equation}
  \label{eq:lfintegral}
  \la {\cal O}[U] \ra_{S_{\rm lat}} \equiv
  \dfrac{\displaystyle\int[DU]{\cal O}[U] e^{-S_{\rm
        lat}[U]}}{\displaystyle\int[DU]e^{-S_{\rm lat}[U]}}
\end{equation}
where $DU$ is the invariant Haar measure. The choice of the lattice
action $S_{\rm lat}$ is quite arbitrary, and 
restricted only by gauge invariance and the requirement that in the
limit of zero lattice spacing it gives back the desired continuum
action. Note that only gauge--invariant operators have non--vanishing
expectation value: this means that the case of parton--parton
scattering, where the relevant operators are the gauge--dependent
Wilson lines, cannot be treated with our approach. 

It is easy to see that in our case the action
\begin{equation}
  \label{eq:mod_E_lat_ac}
  S_{\rm lat} = \beta\sum_{n,\mu < \nu}
  {C}_{E\mu\nu}(\theta,T) \left[1- \frac{1}{N_c}\Re\tr
    U_{\mu\nu}\right],
\end{equation}
where $U_{\mu\nu}$ is the usual plaquette variable (in the
fundamental representation)~\cite{Wil} and $\beta=2N_c/g^2$, gives
back the action $S_E^{\rm Y.M.}$ of Eq.~(\ref{eq:mod_E_ac}) in the
limit $a\to 0$, upon identification of the link variables with
$U_\mu(n)=\exp\{iga\phi_\mu(na)\}$. 
For compact gauge groups, such as $SU(N_c)$, the integration range is
compact, so that, as long as the volume and the lattice spacings are
finite, the integral (\ref{eq:lfintegral}) with the
action (\ref{eq:mod_E_lat_ac}) is convergent and analytic in $\theta$
and $T${ }\footnote{Here we understand that one first restricts to
  $\theta\in{\cal I}_E$ and positive real $T$, and then analytically
  extends ${C}_{E\mu\nu}\to \overline{C}_{E\mu\nu}$.}; in  
the $V\to\infty$ limit, one has to impose positive--definiteness of
the real part of the action in order for the integral to remain
convergent, and this leads exactly to the {\it convergence conditions}
studied in the previous section, with which we have determined the
analyticity domain of ${\cal G}_E$.

The action (\ref{eq:mod_E_lat_ac}) is correct at tree--level, but
one has also to ensure that quantum effects do not modify its
form. Before we discuss this point, it is convenient to recall that
the Euclidean modified action $S_E^{\rm Y.M.}$ has been obtained with independent
rescalings of fields and coordinates in the various directions. 
Indeed, from the definition of $M_E$ in Eq.~(\ref{eq:matrix}) one sees
that the coefficients in the action can 
be written as ${C}_{E\mu\nu}= \lambda_\mu^2\lambda_\nu^2 /
\prod_\alpha |\lambda_\alpha|$ where $\lambda_\mu=M_{E\mu\mu}$, see
Eq.~(\ref{eq:matrix}).  
It is then easy to see that Eq.~(\ref{eq:mod_E_lat_ac}) is also the
correct tree--level action for an anisotropic lattice regularisation
of the usual Euclidean Yang--Mills action, as one can directly check
[see Eqs.~(\ref{eq:E-transf}) and (\ref{eq:matrix})] 
by identifying $U_\mu(n)=\exp\{iga_\mu A_{E\mu}(na)\}$, with
$a_\mu=a/\lambda_\mu$ (note that $\lambda_\mu >0$ for $\theta\in{\cal
  I}_E$ and real positive $T$).
Showing that Eq.~(\ref{eq:mod_E_lat_ac}) is a good lattice action on an
isotropic lattice for the modified action Eq.~(\ref{eq:mod_E_ac}) is
then equivalent to show that it is a good action on an
anisotropic lattice for the usual Yang--Mills action.

As it has been pointed out in~\cite{Bur}, the general anisotropic action 
is not guaranteed to belong to the same universality class as the
isotropic lattice action, and thus one has to enforce that rotation
invariance is restored in the continuum limit to get back the usual
(Euclidean) Yang--Mills action: in the general case one has to
properly tune all the coefficients of the various terms of the action,
obtaining in our case
\begin{equation}
  \label{eq:mod_E_lat_ac_2}
  \widetilde{S}_{\rm lat} = \sum_{n,\mu < \nu}
  \beta_{\mu\nu}{C}_{E\mu\nu}(\theta,T)
  \left[1- \frac{1}{N_c}\Re\tr U_{\mu\nu}\right],
\end{equation}
with properly chosen functions
$\beta_{\mu\nu}=\beta_{\mu\nu}(a,\theta,T)$. 
Due to the
asymptotic freedom property of non--Abelian gauge theories, one can
determine this functions analytically in perturbation
theory for small lattice spacings. One should then check that the
coefficients $\beta_{\mu\nu}$ required to restore rotation invariance
in the continuum limit do not alter the main results derived in the
previous subsections. Quantum effects could in principle impose
further restrictions on the analyticity domain ${\cal D}_E$ found
above, but a preliminary analysis seems to indicate that this is not
the case; however, this issue will be discussed in greater detail in a
separate publication~\cite{moi}. 

We want now to make some remarks on the choice of the operators in the
lattice regularisation of Eq.~(\ref{eq:trans_corrfunc}). After the
field and coordinate transformation, the longitudinal sides of the two continuum
Wilson loops are at $45^{\circ}$ with respect to the new axes, and
have to be approximated by a broken line (see
e.g. Ref.~\cite{lattice}): 
this introduces approximation errors which have to be carefully
considered, but which should vanish 
in the continuum limit, thus leaving unaltered our analysis. To get
rid of this problem, one could use on--axis Wilson loops, thus
performing an ``exact'' calculation on the lattice: to do that one
has to perform a further transformation of the action, choosing the new
basis vectors along the directions of the longitudinal sides of the
loops. The drawback in this case is the appearence of
$\tr[\Phi_{0\alpha_{\perp}}\Phi_{1\alpha_{\perp}}]$ terms
($\alpha_{\perp}=2,3$), which on the lattice correspond to the more
complicated (``chair--like'') terms
$\tr[U_{0\alpha_{\perp}}U_{1\alpha_{\perp}}^\dag]$.

\subsection{Fermions}

The {\it full} (i.e., not {\it quenched}) correlation functions are
obtained including fermion effects in the functional integral via the
fermion--matrix determinant. We can follow the same approach of the
previous subsections also in this case, changing coordinates and 
fields to push the dependence on the relevant variables
$\theta$ and $T$ into
the action: the discussion of analyticity properties can then be made
along the same lines as in the pure--gauge case, representing the
determinant as a functional integral over Grassmann variables and
moving derivatives inside the functional integral. The Grassmannian
integral can always be performed (at least formally), 
resulting in a nonlocal functional of the gauge fields; if we assume
that the Yang--Mills exponential factor is strong enough to ``tame''
this functional, as long as the real part of the exponent is
positive--definite, the procedure goes on exactly as in the pure
Yang--Mills case, and we find the same analyticity structure. Here we
shall limit ourselves to the formal argument, and show that the
analytic--continuation relations and crossing--symmetry relations
remain true also in this case. 

Starting from the Euclidean fermionic action and performing the field
and coordinate transformation (\ref{eq:E-transf}), one
finds the modified fer\-mio\-nic action, which for $\theta\in(0,\pi)$ and
real positive $T$ reads
\begin{align}
  S_{E}^{\rm ferm} &= \left(\frac{T}{T_0}\right)^2\sin\theta\int d^4 y
  \bar{\psi}\left({\cal 
  D}_{\mu}M_{E\mu\nu}\gamma_{E\nu} + m\right)\psi,
\end{align}
where ${\cal D}_\mu \equiv \frac{\de}{\de y^\mu} + ig\phi_\mu$
and $\gamma_{E0} \equiv \gamma^0$, $\gamma_{Ej} \equiv -i\gamma^j$,
with $\gamma^\mu$ the usual Dirac gamma--matrices; the matrix $M_E$ is
given in Eq.~(\ref{eq:matrix}). 
It is then easy to see that the double analytic continuation
$\theta\to -i\chi$, $T\to iT$ actually provides the Minkowskian
fermionic action, and that the crossing--symmetry relations are
reobtained. Indeed, under the analytic continuation  $ {\cal
  D}_{\mu}M_{E\mu\nu}\gamma_{E\nu} \to -i{\cal 
  D}_{\mu}M^{\mu}_{\phantom{\mu}\nu}\gamma^\nu$, so that
\begin{align}
   S_{E}^{\rm ferm} \to & -i\left(\frac{T}{T_0}\right)^2\sinh\chi\int d^4 y \bar{\psi}\left(i{\cal
    D}_{\mu}M^{\mu}_{\phantom{\mu}\nu}\gamma^\nu - m\right)\psi \nonumber\\ 
  = & -iS_{M}^{\rm ferm}, 
\end{align}
where $S_{M}^{\rm ferm}$ is the modified Minkwskian action, obtained
performing the transformation of fields and coordinates in Minkowski
space--time [see Eqs.~(\ref{eq:M-transf}) and (\ref{eq:matrix})],
as expected.

The exchanges $\theta\to\pi-\theta$ and $\chi\to i\pi-\chi$ in the
Euclidean and Minkowskian theories, respectively, are equivalent
to $\phi_0 \leftrightarrow \phi_1$, $y^0\leftrightarrow y^1$, provided
one also performs the following change of variables in the
Grassmannian integration, 
\begin{equation}
  \left\{
  \begin{aligned}
    & \psi \to U\psi\\
    & \bar{\psi} \to \bar{\psi}U^\dag
  \end{aligned}\right.
\end{equation}
with
\begin{equation}
  U = \gamma_{E5}\frac{\gamma_{E0} -\gamma_{E1}}{\sqrt{2}} =
    \gamma^5\frac{\gamma^0 +i\gamma^1}{\sqrt{2}},
\end{equation}
where
$\gamma_{E5}=\gamma_{E0}\gamma_{E1}\gamma_{E2}\gamma_{E3}=i\gamma^{0}\gamma^{1}\gamma^{2}\gamma^{3}=\gamma^5$,
in order to exchange also the longitudinal gamma matrices,
\begin{equation}
  \left\{
  \begin{aligned}
    U^\dag \gamma_{E0} U &= \gamma_{E1}, &     U^\dag \gamma^0 U &= -i\gamma^1\\
    U^\dag \gamma_{E1} U &= \gamma_{E0}, &     U^\dag \gamma^1 U &= i\gamma^0.
  \end{aligned}\right.
\end{equation}
Note that $U$ is antihermitian and unitary, $U^\dag = -U = U^{-1}$.
In this way the exchanges $\theta\to\pi-\theta$ and $\chi\to
i\pi-\chi$ are seen to be equivalent to the exchange of one of the two
loops with the corresponding antiloop, thus extending the validity of
the crossing--symmetry relations~(\ref{eq:crossrel}) to the case where
also fermions are included. One can also easily show, combining
Eq.~(\ref{eq:refl0}) and $C$--invariance, that the {\it reflection relation}
(\ref{eq:reflG}) still holds after the inclusion of fermions.

\section{Concluding remarks and prospects}

In this letter we have approached the analyticity issues related to
the problem of {\it soft} high--energy scattering by means of
functional--integral techniques, giving a nonperturbative justification of the 
hypotheses underlying the analytic--continuation relations between the
relevant Wilson--loop correlation functions in the Euclidean and
Minkowskian theories. The argument relies on a transformation of
coordinates and fields that moves the whole dependence on the relevant
variables, namely the angle $\theta$ between the loops and the
half--length $T$, into a modified action; then, the {\it convergence 
conditions} on the functional integral give rise to a nontrivial
analyticity domain for the Euclidean correlation function, which is
sufficiently wide for the analytic--continuation relations, and also
for the crossing--symmetry relations, obtained
in~\cite{Meggiolaro97,Meggiolaro98,Meggiolaro02,Meggiolaro05,crossing,Meggiolaro07}, 
to hold; moreover, these relations are reobtained in a completely 
independent way. 

To put the argument on a more solid ground, we have employed a lattice
regularisation of the functional integral. Here analyticity of the
correlation function follows from the compactness of the integration
range as long as the volume and the lattice spacings are finite, and
the {\it convergence conditions} are necessary conditions for the
convergence of the Haar integral in the limit of infinite volume. To
ensure the correct continuum limit, the tree--level action should be
corrected taking into account quantum effects: in principle this could
lead to further restrictions on the domain of analyticity, but a
preliminary analysis seems to indicate that this is not the case.
This issue will be investigated in greater detail in a separate
publication~\cite{moi}. 
Also, the infinite--volume limit and the zero--lattice--spacing limit
can be sources of singularities if the 
convergence is not uniform: to prove that this does or does not happen
is a very hard problem, which we have not attempted to tackle here.
Singularities could also appear if the Wilson--loop expectation value
vanishes at some complex value of $\theta$ and $T$: while poles are
not a problem for the analytic continuation, the presence of algebraic
singularities can cause an ambiguity in the choice of the Riemann
sheet. 

One can be tempted to take the large--$T$ limit directly in the
action, or to perform the analytic continuation from Euclidean to
Minkowski space and then take the large--$\chi$ (i.e., high--energy)
limit. These limits have to be taken very carefully: for example, if
one keeps only the leading order in $T$ (or in $\chi$) in the action
of the lattice--regulated functional integral, one obtains exactly
zero for both the (unnormalised) correlation function and the
Wilson--loop expectation value, leaving the correlation function
${\cal G}_E$ undetermined. 

The modified Euclidean action derived in section 3 can however
be used as a starting point for a nonperturbative investigation 
of {\it soft} high--energy scattering from the first principles of QCD, in
principle also from a numerical point of view. This is similar to the
approach adopted in Refs.~\cite{Verlinde,Arefeva,Orland}, where other
rescaled actions have been proposed. This issue (including a detailed
study of quantum corrections) will be investigated in another
publication~\cite{moi}.

\pagebreak

\end{document}